# Dynamics of flare processes and variety of the fine structure of solar radio emission over a wide frequency range of 30 - 7000 MHz

Gennady Chernov [1,2], Valery Fomichev [2], Baolin Tan [1], Yihua Yan [1], Chengming Tan [1], Qijun Fu [1]

[1] *Key Laboratory of Solar Activity, National Astronomical Observatories Chinese Academy of Sciences, Beijing 100012, China*
[2] *Pushkov Institute of Terrestrial Magnetism, Ionosphere and Radio Wave Propagation, Russian Academy of Sciences (IZMIRAN), Moscow, Troitsk, 142190, Russia*

**Abstract** Radiobursts exibiting fine structure observed over two years during the rising phase of Cycle 24 by the Chinese Solar Broadband Radio Spectrometer (SBRS/Huairou) in the range 1–7.6 GHz and the spectrograph IZMIRAN in the meter range (25 - 270 MHz) are analyzed. In five events zebra structure, various fiber bursts and fast pulsations were observed. These observations have great importance for testing different theoretical models of fine structure formation, as, for example, only for explaining the zebra-structure more than ten mechanisms have been proposed. Events on 15 and 24 February 2011 are of the greatest interest. In the course of the flare on 15 February (which occurred close to the center of disk) zebra structure was observed during three sequential flare brightenings. The polarization changed sign in the third. This behavior of polarization combined with images of the corresponding flare brightenings, obtained in extreme ultraviolet radiation by the *Solar Dynamics Observatory* (SDO/AIA, 171 Å) provides important clues. The polarization of radio emission in all three cases is related to the ordinary wave mode of radio emission. The zebra structure was present at frequencies 190-220 MHz in the Culgoora spectrum. The event on 24 February 2011 is remarkable, as the zebra structure at frequencies of 2.6-3.8 GHz was not polarized and it appeared during the magnetic reconnection observed by SDO/AIA 171 Å in this limb flare. In the event on 9 August 2011, for the first time, a superfine millisecond structure was registered simultaneously in the fast pulsations and the stripes of zebra structure. In the event on 1 August 2010 after the zebra structure two families of fibers bursts with an opposite frequency drift were observed. On 19 April 2012 the fibers against the background of type III bursts were observed by IZMIRAN and Nançay spectrographs. In the band of 42 – 52 MHz a group of nine slowly drifting narrow-band (</~ 1 MHz) fibers formed a unique type II burst. Almost all events in the microwave range contain superfine structure in the form of the millisecond spikes, whose emission should be considered primary. It is possible that each type of fine structure is excited by the same mechanism, and the broad variety of events is related to the dynamics of flare processes.

*Subject headings:* Sun: flares − Sun: fine structure − Sun: microwave radiation

## 1. INTRODUCTION

Solar radio astronomy has grown into an extensive scientific branch since its birth in the 1940s, initiated by the subsequent discovery of the main basic components of solar radio emission: the quiet Sun, the slowly varying component and various types of radio bursts including noise storms (Krüger, 1979). It was quickly revealed, that radio bursts are observed during chromospheric flares. Already the first spectral observations of large type IV (and II+IV) bursts revealed the rich variety of the fine structure of the radio emission in the form of wide-band pulsations in emission and absorption with the different periods, rapid bursts (spikes), narrow-band patches. Modulation of the continuum emission in the form of narrow stripes in the emission and the absorption in dynamic spectra (zebra pattern) has been one of the most intriguing elements of the fine structure (Krüger, 1979). Zebra pattern forms as a manifestation of special plasma processes in the corona. Therefore they are of special interest not only for solar physics but also for plasma physics.

The nature of the zebra pattern (ZP) has been a subject of intense discussion for more than 40 years. The ZP in the solar radio emission is the simultaneous excitation of waves at many (up to a few tens) of closely spaced, nearly equidistant frequencies. The basic parameters of ZP in the meter wave band are represented in the atlas by Slottje (1981). Now it is well established that in a regular ZP the frequency separation between the stripes grows with frequency: from 4 – 5 MHz at 200 MHz to ~ 80 MHz at 3000 MHz and to ~150 – 200 MHz at 5700 MHz. It is important to note that the relative frequency bandwidth of a separate stripe in emission remains almost stably constant with frequency, $\Delta f_e/f \approx 0.005$.

More than ten different models have been proposed for ZPs; most of them include some emission of electrostatic plasma waves at the upper hybrid frequency ($\omega_{UH}$) (Kuijpers, 1975a; Zheleznykov and Zlotnik, 1975a,b; Mollwo, 1983; 1988; Winglee and Dulk, 1986). The most comprehensively developed ZP models involve mechanisms based on the double plasma resonance (DPR), which assumes that the upper hybrid frequency in the solar corona becomes a multiple of the electron-cyclotron frequency:

$$\omega_{UH} = (\omega^2_{Pe} + \omega^2_{Be})^{1/2} = s\omega_{Be} \qquad (1)$$

where $\omega_{Pe}$ is the electron plasma frequency, $\omega_{Be}$ is the electron cyclotron frequency, and $s$ is the integer harmonic number.

In order to explain the ZP dynamics in the framework of this mechanism, it is necessary for the magnetic field in the radio source to vary sufficiently rapidly, which, however, contradicts the fairly low field values determined from the frequency separation between the stripes. Over the past five years, dozens of articles were published aiming at the refinement of this mechanism, because, in its initial formulation, it failed to describe many features of the ZP. Kuznetsov and Tsap (2007) assumed that the velocity distribution function of hot electrons within the loss cone can be described by a power law with an exponent of 8–10. In this case, a fairly deep modulation can be achieved, but the excitation of waves at multiple double plasma resonance (DPR) levels is still impossible.

Fiber bursts differ from ZP stripes only by a constant negative frequency drift, and one of the first models explained the radio emission ($t$) of fiber bursts by the coalescence of plasma waves ($l$) with whistlers ($w$), $l + w \rightarrow t$ (Kuijpers, 1975a). In spite of the fact that then some more models were proposed (Alfvénic solitons, whistler solitons, MHD oscillations), this first model remains the most widely accepted. After Kuijper's review (Kuijper, 1980), the topic was revisited only in the review of Chernov (2006).

In Chernov (1976; 1990), the mechanism $l + w \rightarrow t$ was proposed as a unified model in which the formation of ZPs in the emission and absorption spectra was attributed to the oblique propagation of whistlers, while the formation of stripes with a stable negative frequency drift (the fiber bursts) was explained by the ducted propagation of waves along a magnetic trap. This model explains occasionally observed transformation of the ZP stripes into fibers and vice versa.

The discovery of the superfine structure of the ZP, in the form of millisecond spikes was the most significant new effect in the microwave range (Chernov, Yan and Fu, 2003). The reliability of such a study strongly improved over the past few years due to numerous observations of fast radio bursts (millisecond spikes) during the stellar flares (Abada-Simon et al., 1995). It is amazing that the period of stellar spikes in the radio burst of the classical red dwarf AD Leo nearly the same as the period of spikes in the superfine structure of solar zebra-stripes (~30 ms) (Osten and Bastian, 2006).

To overcome difficulties arising in different models, a new ZP theory based on the emission of auroral choruses (magnetospheric bursts) via the escape of the Z mode captured by regular plasma density inhomogeneities was proposed (LaBelle et al. 2003). This theory, however, fails to explain the high intensity of radiation emitted by separate incoherent sources. In addition, the theory imposes some stringent conditions, such as the presence of a large amplitude ion–acoustic wave.

The existence of a ZP in the solar radio emission can be attributed to the existence of discrete eigenmodes in the nonuniform solar atmosphere. Several aspects of this mechanism were considered in Laptukhov and Chernov (2006); Bárta, M. and Karlický (2006); Ledenev, Yan, and Fu (2006). In Laptukhov and Chernov (2006), dispersion relations were derived for a discrete spectrum of eigenmodes of a spatially periodic medium in the form of nonlinear structures formed due to the onset of thermal instability. The spectrum of eigenfrequencies of a system of spatially periodic cavities is calculated, and it is shown that such a system is capable of generating a few tens of ZP stripes, the number of which is independent of the ratio of the plasma frequency to the gyrofrequency in the source.

In practically all models the discussion deals with regular ZP. Problems appear with the interpretation of the frequently observed uncommon stripes of a ZP. For example, for explaining the so-called "tadpoles" (submerged in a developed ZP) special mechanisms were elaborated (Zheleznykov and Zlotnik, 1975a; Chernov, 2006). Rare attempts were made to explain other uncommon forms of ZPs (in the form of zigzags, complex splittings of stripes into the superfine structures). They cause big problems for the existing models.

It is important to note that the whistler model successfully explains the zigzags of stripes and their splitting, and also the variations in the frequency drift of stripes synchronously with the spatial drift of the sources of radio emission (Chernov, 2006). Since each new phenomenon provides its specific parameters of fine structure, and this variety of the parameters does not yield to the statistical analysis, below primary attention is given to the analysis of separate phenomena. Just such a situation stimulates many authors to elaborate on new mechanisms.

In the present paper, an attempt is made to evaluate what model most adequately describes the new observational data and to find out where the ZP stripes form (during the excitation of waves in the source or in the course of their further propagation). Calculations show that the DPR-based mechanism fails to describe the generation of a large number of ZP stripes in any coronal plasma model. Some other unsolved problems or difficulties in the DPR model are also examined in detail. Here, it is shown that the new varieties of ZP succeed in explaining these phenomena within the framework of known mechanisms by taking into account the special features of plasma parameters and fast particles in the source. On the other hand, the formation of ZP stripes due to radio wave propagation through the coronal heterogeneities can be recognized as the most natural mechanism of ZP. The mechanism related to the excitation of discrete eigenmodes of the periodically nonuniform plasma (Laptukhov and Chernov, 2006, 2009) can yield the observed number of harmonics. However, in this case, only the possibility of generating harmonics in a one-dimensional stationary case is considered, i.e., the frequency dynamics of stripes is not analyzed.

During the last several years some new types of ZP have been recorded. Now, it is necessary to consider the possibility of their interpretation by taking into account all known models of ZPs.

## 2. OBSERVATIONS

### 2.1. 24 February 2011 event

The radio burst was of small duration (~8 minutes), and the first 7 minutes it displayed smooth growth and fall with a number of irregular pulsations in the range 2.6 - 3.8 GHz (Figure 1). A rich fine structure appeared in the second more powerful maximum during approximately 30 sec after 07:37 UT (Figure 2).

The position of the corresponding flare on the eastern limb caused special interest in this event. The flare of the importance M3.3 occurred in the active region NOAA 11163 with the coordinates N14 E87. This position of flare made possible to see real advance into the corona of two ejections with an interval of ~7 min, which were observed in the extreme ultraviolet lines on board of SDO/AIA and of the subsequent coronal mass ejection, which was observed in the white light with the coronograph SOHO/LASCO C2.

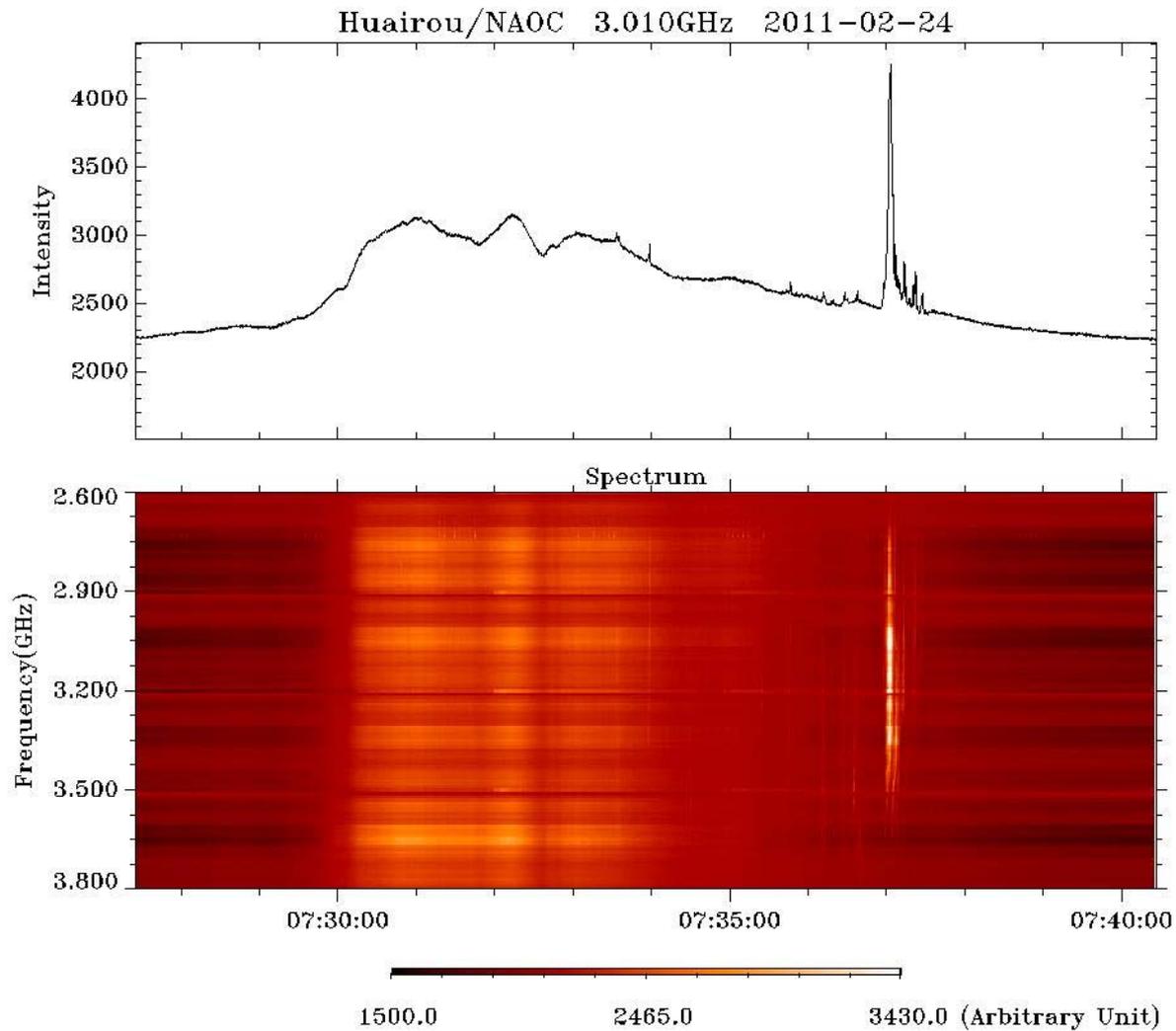

Figure 1. General view of the radio burst at the fixed frequency of 3.010 GHz and the dynamical spectrum in the frequency band of 2.6 – 3.8 GHz observed by the SBRS/Huairou of the 24 February 2011 event.

These data allow to understand the dynamics of flare processes, in particular, why the fine structure of radio emission was observed only during the second short flare brightening. In the meter range type II burst was observed, and these optical data support the idea that the corresponding shock wave was of piston type originating from the first ejection.

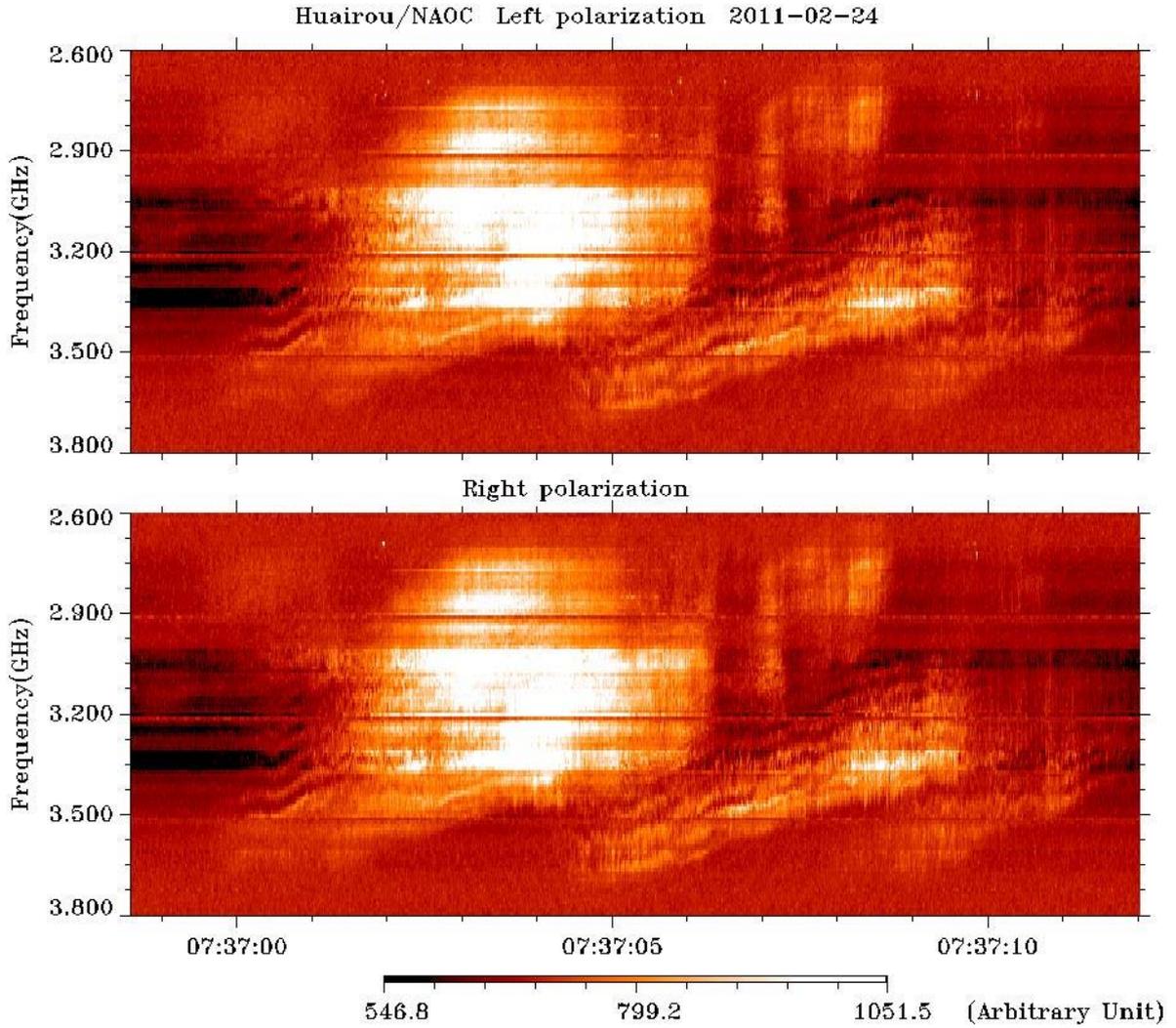

Figure 2a. Dynamical spectra of zebra structure and pulsation during 13 sec registered by the SBRS/Huairou after the second flare ejection. The numerous stripes of zebra-structure reveal the superfine structure in the form of millisecond spikes. The radio emission is not polarized.

In Figure 2a it is obvious that the second burst consists of the stripes of zebra-structure (ZS) which are wavy drifting with an average speed of $df/dt \sim -60$ MHz/sec. On low-frequency (LF) edge of the range the zebra structure disappears, and then pulsations with the irregular period are visually allocated.

The frequency separation between the stripes of ZS decreases with the decrease of frequency from 75 MHz on 3600 MHz to $\sim 40$ MHz on 3000 MHz. On the high-frequency (HF) edge of the range all the emission consists from spikes with the duration of $\sim 30$ ms. These are usual parameters of ZS in the microwave range, according to Chernov (2011). This superfine structure almost disappears on the LF edge of the range. In the Left and Right channels of spectropolarimeter the intensity of emission is practically identical, i.e. radio emission is not polarized.

During the decay of the first prolonged burst we can see several weak peaks (Figure 1). In one of them, 1.5 min prior to the second burst it is possible to see the weakly expressed small-scale zebra-structure (frequency separation of $\sim 20$ MHz) (Figure 2b) almost in entire range 2.6 - 3.8 GHz. However, the strictly constant intermediate drift of the stripes $df/dt$ of $\approx$ of $-300$ MHz/s allows to carry them rather to the fiber bursts.

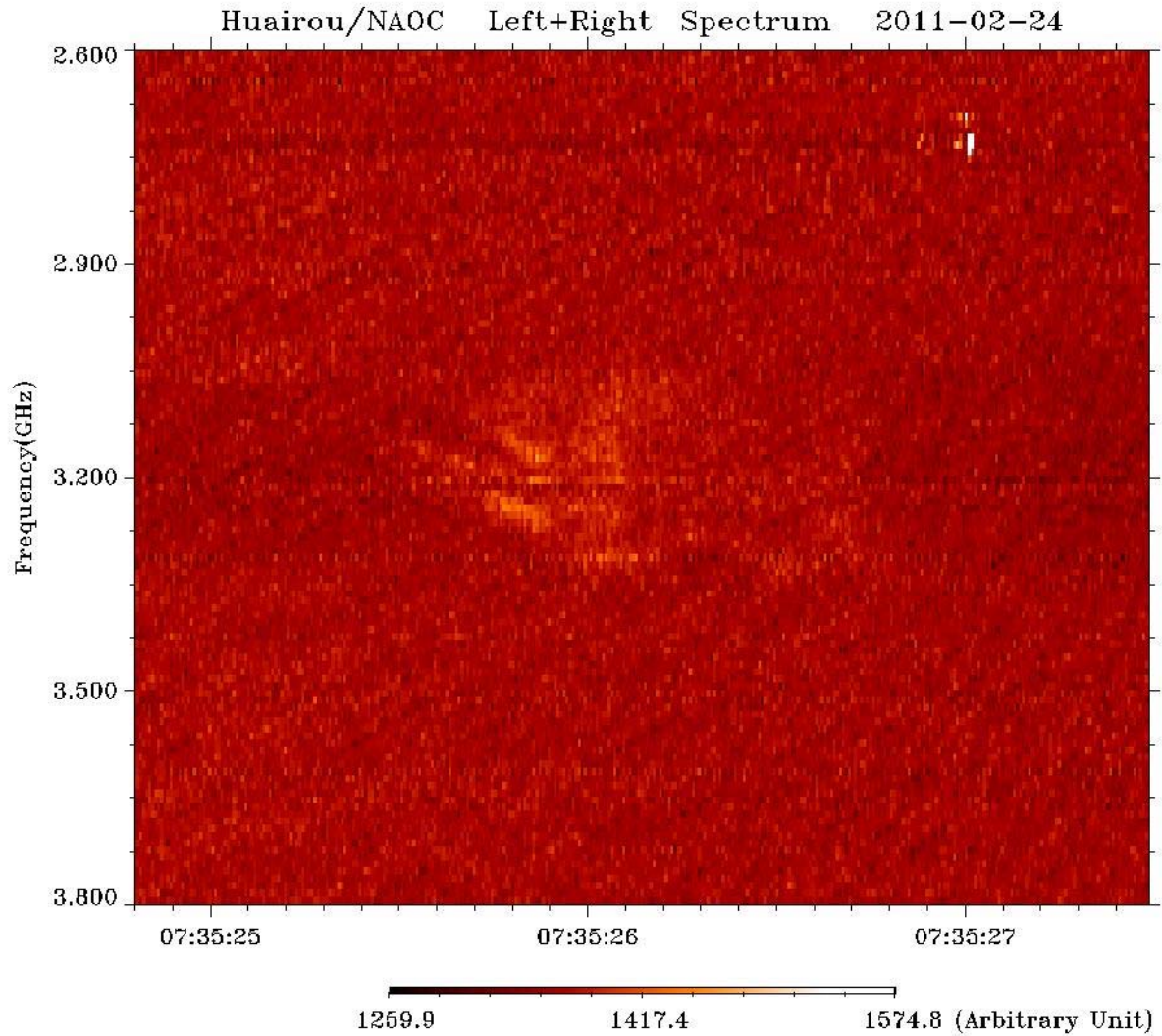

Figure 2b. A faint small-scale zebra structure about 1.5 min prior to the second burst.

The beginning of the first ejection at 07:30:24 UT and its continuation (at the moment of the maximum of the first prolonged burst at 07:32:02 UT) are shown in Figure 3 by two SDO/AIA 171 Å images. The beginning of the ejection has a form of a twisted magnetic flux-rope.

The II burst (Figure 4) that consists of two harmonics, according to the Hiraiso spectrograph (Japan), starts at 07:35 UT when the most dense part of the first emission reaches about ~ 150 000 km height (Figure 3) suggesting a piston-driven shock wave.

In Figure 5 the first harmonic of the II type burst observed at IZMIRAN is shown. The burst spectrum has a complex, ragged structure, indicating strong heterogeneity behind the shock wave front.

Figure 6 presents two frames from the SDO/AIA 175Å movie during the second ejection coinciding with the second maximum of radio burst whose fine structure in the microwave range is shown in Figure 2a.

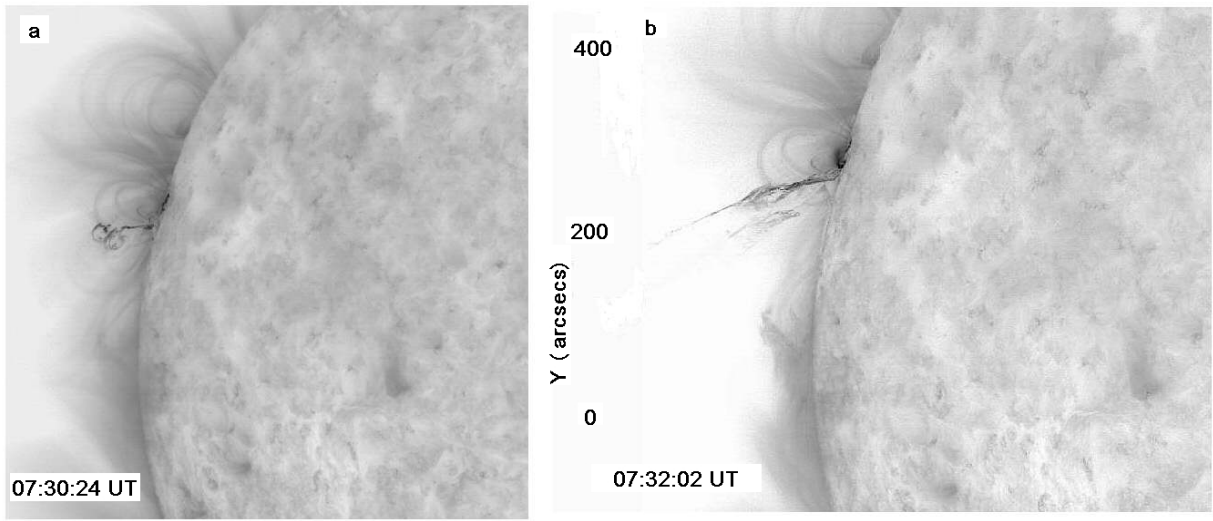

Figure 3. Two frames from the movie of SDO/AIA 171 Å showing the beginning of the first ejection at 07:30 UT (a) and its continuation at 07:32 UT (b).

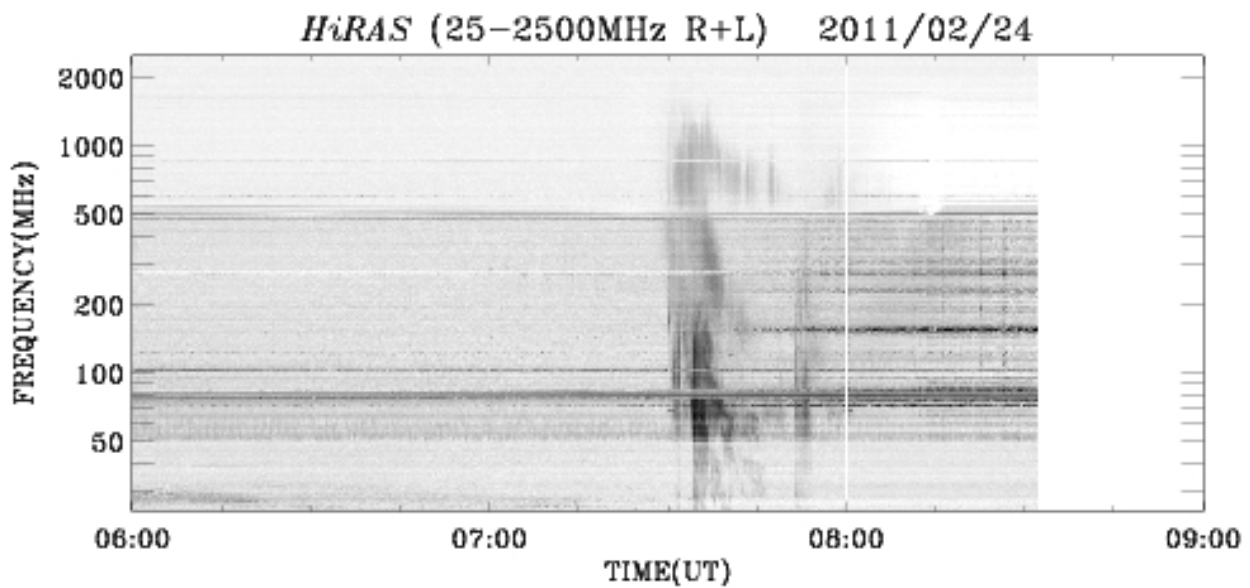

Figure 4. The radio burst observed by the Hiraiso spectrograph (Japan) in the range of 25 - 2000 MHz. The burst beginning in the form of several groups of the type III emissions coincides with the beginning of the first ejection at ~ 07:30 UT.

The beginning of the second ejection at 07:37: 02 UT (Figure 6a) also has a shape of a twisted magnetic rope. The frame showing the end of this ejection at 07:38: 48 UT (Figure 6b) is of the greatest interest. It is evident that the magnetic rope extended high in the corona. But its lower part remains approximately at the same height above the point-like brightening, similar to an X- point of magnetic reconnection, and bright lower loops slightly disperse closer to the flare region. The height of this X- point occurs about of 45000 km; therefore it is possible that the microwave emission of fine structure must escape from lower altitudes, from the divergent bases of loops.

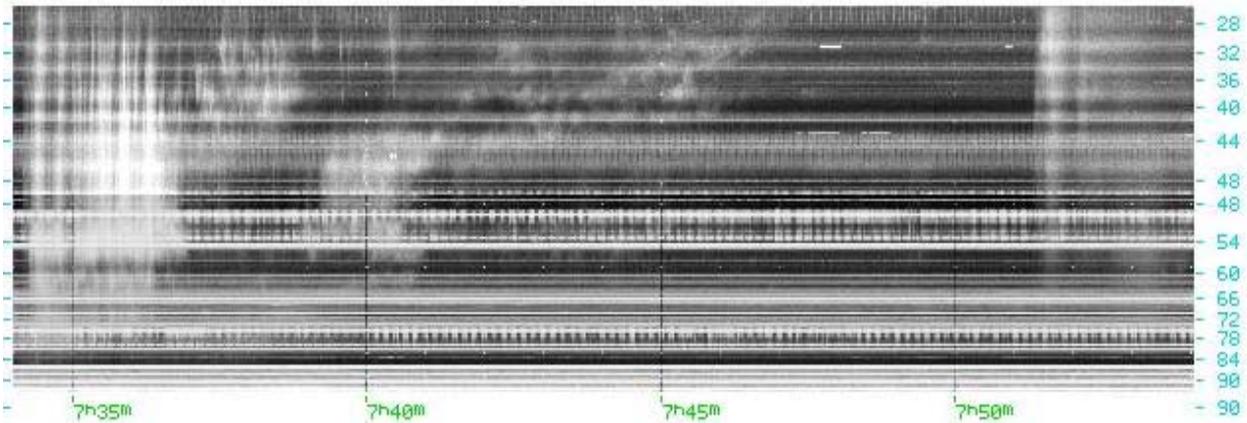

Figure 5. Type II burst which was observed with the IZMIRAN spectrograph in the frequency band 25 – 90 MHz on 24 February 2011.

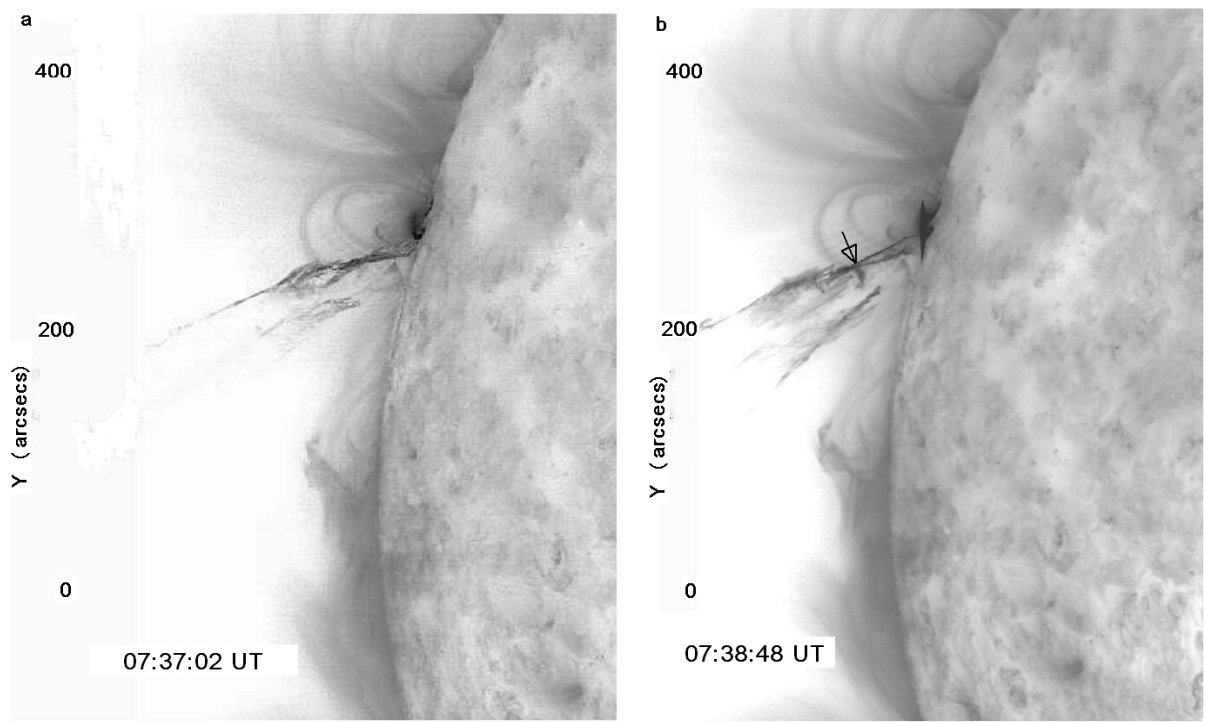

Figure 6. Two frames from the movie of SDO/AIA 171 Å showing the beginning of the second ejection at 07:37:02 UT (a) and its continuation at 07:38:48 UT (b). In the frame b the ejection has form of a magnetic reconnection with an X- point (see arrow). The thickness of the loop bases under the X- point is of ≤ 1".

In the course of this second maximum of radio burst there were five separate peaks (see Figure 1), and they coincide on time with five flare brightenings, registered on AIA 171Å.

Thus, we for the first time observe radio burst with fine structure in the microwave range from the bases of loops under an X- point of magnetic reconnection. These bases of loops are very thin, having a width of less than 1".

As the first ejection extended high into the corona, there was no magnetic trap for fast particles; therefore a continuum in the meter range, in which the fine structure is usually formed was not observed. The coronal mass ejection, which appeared in LASCO C2 after 07:36 UT (Figure 7) was a continuation of this ejection. The second ejection has risen above the X- point of magnetic reconnection. The particles, accelerated upward in the course of magnetic reconnection, led to a group of type III bursts (or pulsations) at frequencies 800 - 500 MHz (see Figure 4).

The particles, accelerated downward, were captured into a magnetic trap, or more precisely, they were responsible for the formation of fine structure in the microwave range, zebra-structure and pulsations, shown in Figure 2.

It is natural to connect zero polarization with the depolarization of the radio emission, which propagated perpendicularly to magnetic field from the limb source.

One more feature of this event is represented by incidental emergence of the zebra-structure more than 4 minutes later the second maximum in very low level of a continuum (Figure 8). It is possible to assume that the magnetic trap under the reconnection area existed for a long time.

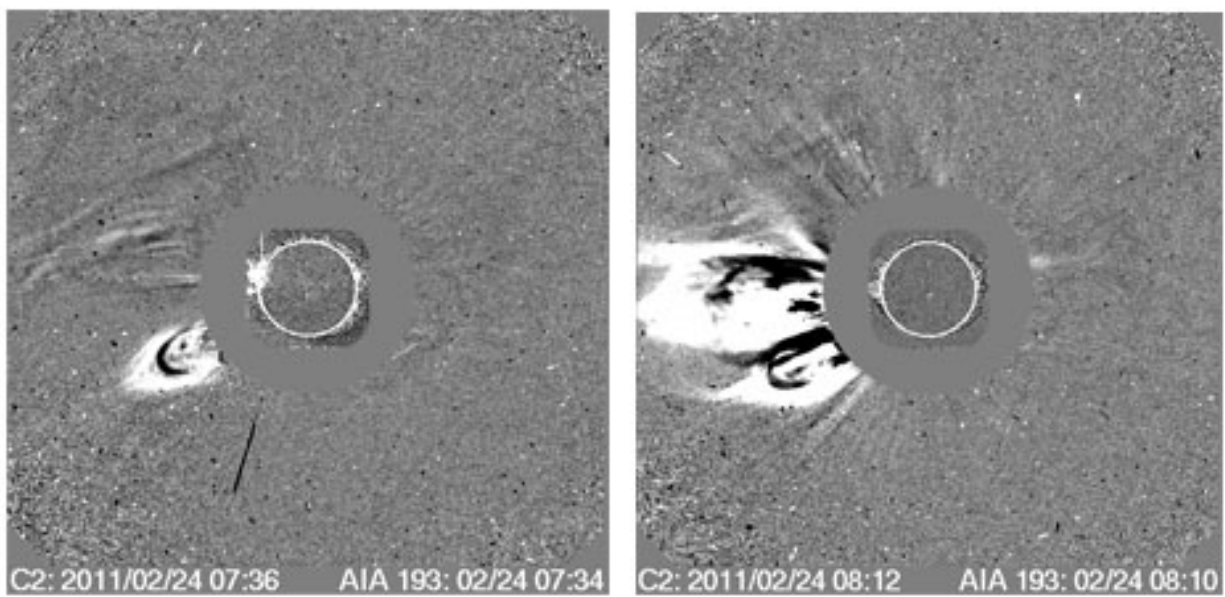

Figure 7. Difference images of the CME, caused by the first ejection. The loop-like CME to the south originated from the previous ejection approximately from 02:00 UT.

In Figure 8 it is possible also to notice the superimposition of zebra stripes with different scales of frequency separation. Possibly, the different scales of stripes are related with simultaneous formation of zebra-structure at the two footpoints of the trap having different parameters.

Thus, in this event we, for the first time, registered the appearance of a zebra-structure at the moment of magnetic reconnection high in the corona. Furthermore, appearance of weak ZS at the very low level of continuum is the new important fact, which requires separate examination.

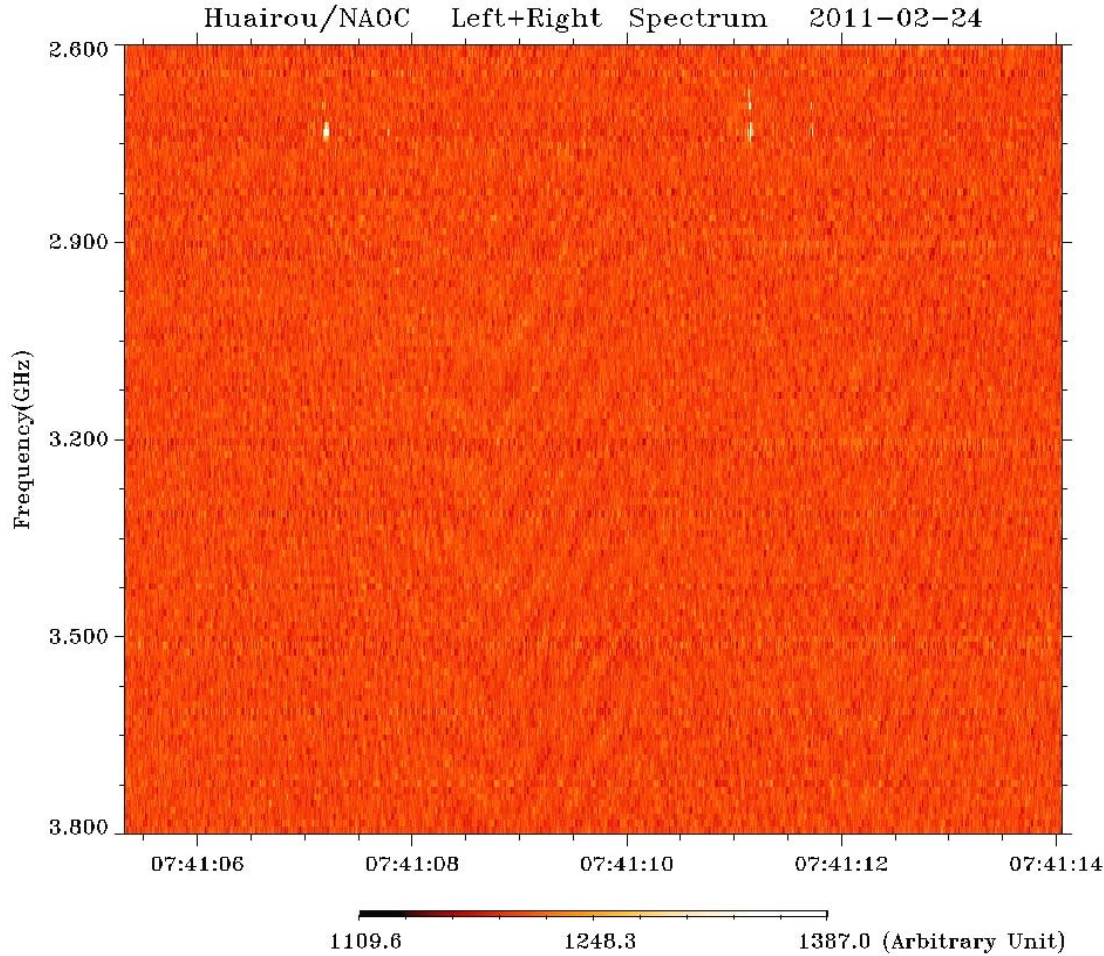

Figure 8. Faint zebra-structure about 4 min after the second burst superposed on a very low level of the continuum.

## 2.2. 15 February 2011 event

The flare on 15 February 2011 (01: 48 - 02:30 UT) was the first X-class flare (2.2) in Cycle 24. It occurred close to the center of disk (S21 W21) in NOAA AR 11158. Many aspects of this phenomenon have already been discussed in a series of the works, mainly dedicated to acoustic oscillations and seismology. Tan et al (2012) examined three intervals with the zebra- structure in different frequency ranges. Besides radio-spectral observations with the SBRS in Huairou (Peking and Yunnan) the data of the *radio polarimeter of Nobeyama* (NoRP), *Nobeyama radio heliograph* (NoRH) and data of SDO were used. Figure 1 in Tan et al (2012) depicts the temporary profiles of the fluxes of radio emission in a number of frequencies and the profiles of soft X-ray GOES 4 and 8. The X-ray flare began approximately 15 min earlier than radio bursts. The temperature of plasma in the X-ray sources was very high, on the order of 20 MK (see panel 9 in Figure 1 in Tan et al (2012)).

Three ZS stripes in the microwave range appeared first at frequencies 6.4 - 7.0 GHz in the left handed polarization during the rising phase of the first flare brightening (see ZP1 in Fug.2 in Tan et al. (2012)). Then, in ~ 11 min ZP2 appeared on the decrease of the second brightening in the range 2.6 - 2.75 GHz (shown here in Figure 9) also in the left handed polarization. Further, still in ~ 9.5 min on the decrease of the third brightening in the decimeter range 1035 - 1050 MHz, ZP3 was registered by the SBRS/Yunnan (see Figure 5 in Tan et al. (2012)), but already with the moderate right-handed polarization. Thus, in the prolonged event (~ 45 min) the ZS appeared only three times by three-second intervals.

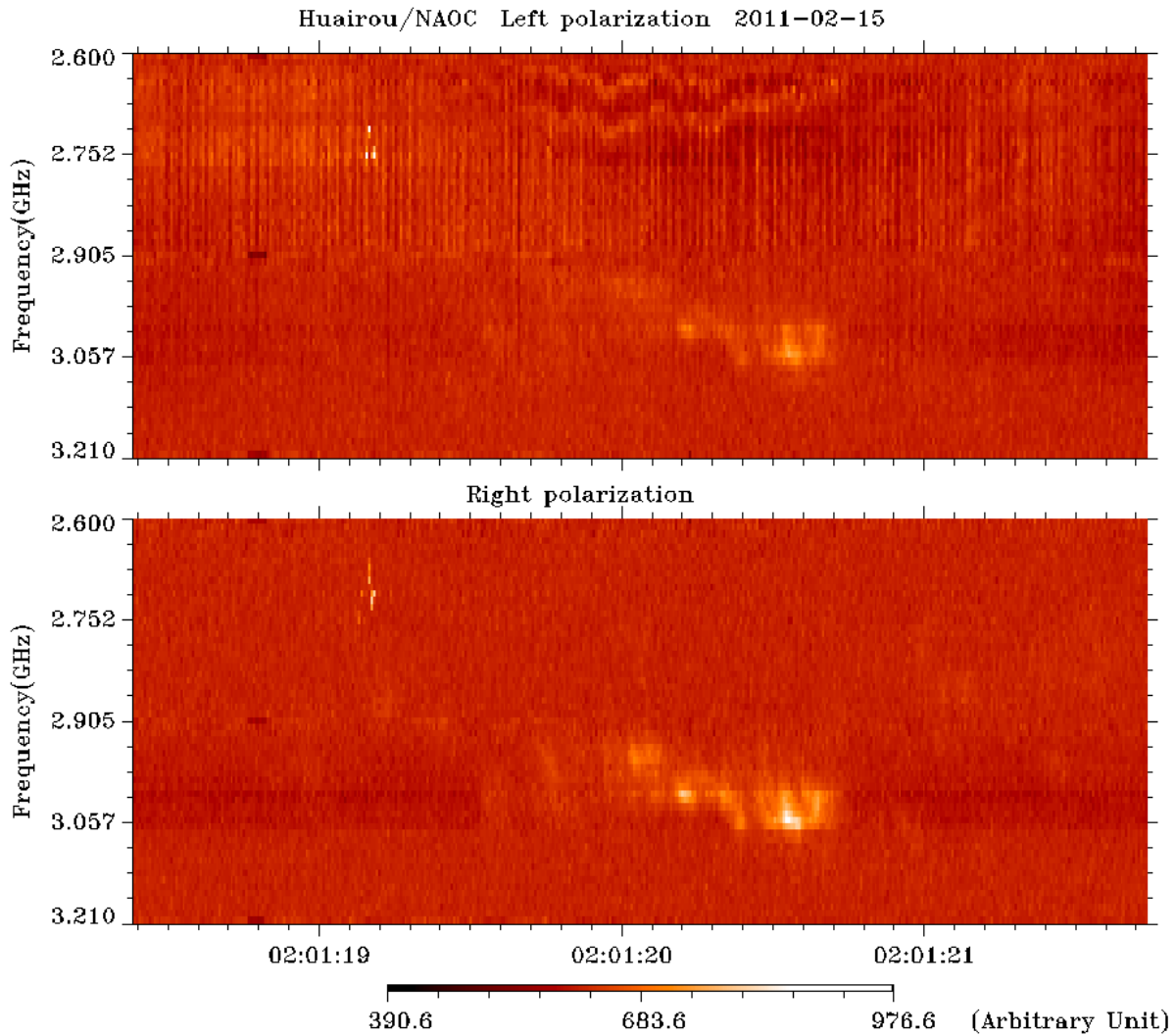

Figure 9. Two stripes of ZS superimposed on fast pulsations with a period of ~ 30 ms registered by the SBRS/Huairou.

Here, we focus on a number of aspects, not examined in the work of Tan et al. (2012). In particular, zebra-structure was observed also in the meter range, with the *Culgoora spectrograph*. In Figure 10 it is possible to distinguish the zebra-structure in the range 200 - 300 MHz. The type II burst is also visible, which begun at 01:50 UT. The onset of the corresponding CME was approximately at 01:30 UT, that coincides with the beginning of the X-ray burst. Therefore the type II burst was most likely caused by a piston shock wave.

Besides, in Tan et al. (2012) the superfine structure of stripes in the form of millisecond spikes (Figure 9) was not considered. In this range it is also seen that coincidentally with two stripes of the ZS in the left channel, strong ragged radio bursts appeared in the right channel. Stripes of the ZS are embedded in a series of the fast pulsations in the frequency range of 2.6 - 2.9 GHz with the period of ≈ 30 ms. After approximately 2.5 min. specifically, in this range fiber bursts appeared with the intermediate frequency drift of *df/dt* of ≈ - 350 MHz s$^{-1}$, approximately with the same period (Figure 11). Subsequently, during five minutes the initial frequencies of fibers increased approximately on 100 MHz together with a new series of pulsations with an irregular period.

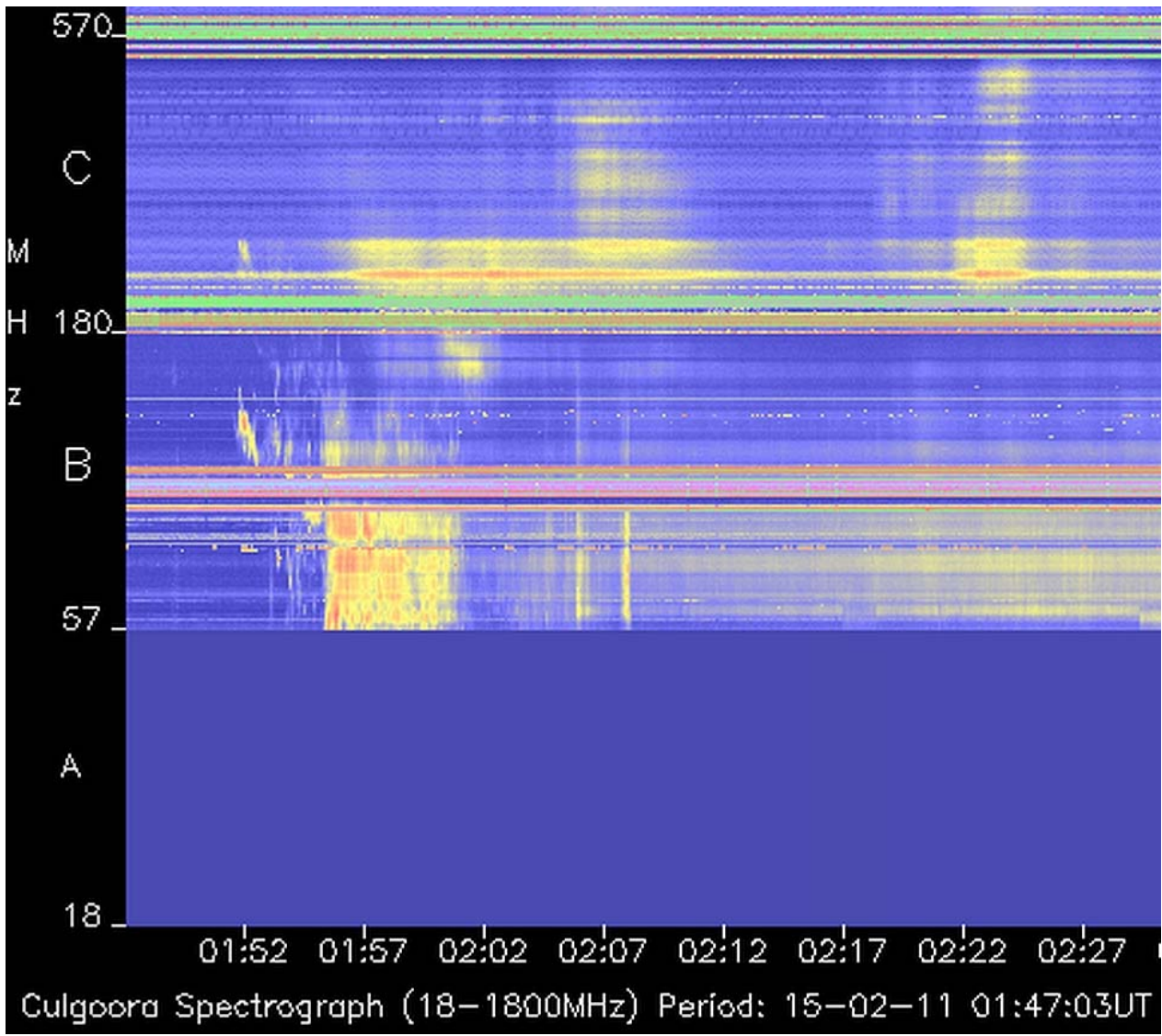

Figure 10. The Culgoora spectrum of 15 February 2011 event with zebra-structure in the range 200 - 300 MHz.

If the onsets of three intervals ZP1, ZP2 and ZP3 are related to the appropriate flare brightenings and the ejections observed in the SDO/AIA of 171 Å images, then they originate in the different sections of AR. In Figure 12 it is shown that ZP1 and ZP2 comes from the loops above the preceding spot of positive magnetic polarity, and ZP3 above the trailing spot of negative polarity. Therefore in all three cases the radio emission can be connected with the ordinary wave mode.

## 2.3. Other events

Figures 13 – 16 show entire variety of appearance of the ZS and fiber bursts in different events. In the 1 August 2010 event (Figure 13) two families of fibers on the LF edge of the range with opposite frequency drift appeared simultaneously with a fragment of ZS at higher frequencies. But in the meter range in this event only irregular pulsations were observed, although the level of continuum was sufficiently high (Figure 14). In Figure 15 the emergence of ZS in LF edge of the range of 2.6 - 3.15 GHz (imposed on fast pulsations) is shown. The new effect here consisted of for the first time the superfine millisecond structure was observed not only in ZS, but also in pulsations. This event was in details discussed in the work of Tan and Tan (2012).

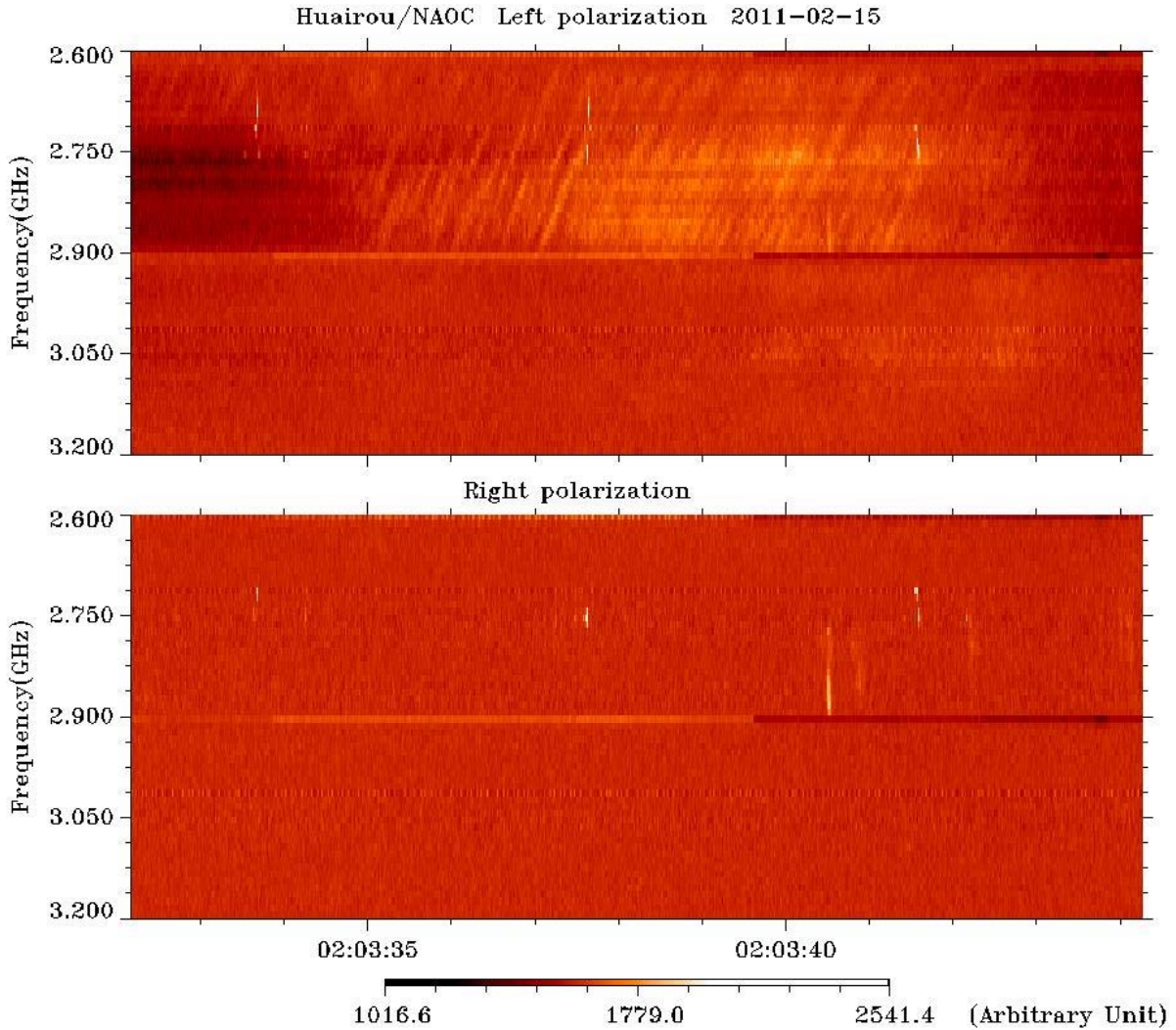

Figure 11. Fibers with the intermediate frequency drift appeared in the same range as millisecond pulsations in Figure 9.

In Figure 16a the unusual type II burst consisting of fibers in the frequency range of 42 - 52 MHz is shown. A stronger emission of type III bursts was superimposed on fibers. The narrow instantaneous frequency bandwidth of each fiber (</~ 1 MHz) allows to classify them as fiber bursts although they are not parallel with each other.

Figure 19b shows the spectrum of this phenomenon received with *NANCAY Decametric Array* in the range 10–80 MHz. All large details coincide with IZMIRAN spectrum. But here, more sensitive tool finds other family of weak fibers with a reverse frequency drift and whole ensemble of point-like spikes.

## 3. DISCUSSION

The events examined here show that the zebra-structure and fiber bursts can appear almost simultaneously or consecutively in the microwave, decimeter and meter wave bands. In the 1 August 2010 event they were almost superimposed on each other in the range 3.2 - 2.6 GHz. Certainly this fact was known earlier. For example, in the meter range see Figure 4.9 in Chernov, 2011 and in the microwave – ibid, Figure 4.38. However when interpreting of such phenomena authors usually artificially mark out two idealized structures: parallel drifting (probably wavy) stripes of the zebra- structure and fiber bursts with intermediate frequency drift (constant and negative). Though the phenomena in which stripes of the zebra structure smoothly pass to fibers and back are known (see, for example, in monographs Chernov (2011) Figure 4.13 and 4.22). Figure 5 in Tan et al. (2012) can also serve as an example of a such passage of one structure into another (according to their determination).

In the theory of formation of these structures most often any smooth transition of one structure to another it isn't considered at all. In the theory of zebra- structure the wide acceptance obtained the mechanism of double plasma resonance (DPR), when the upper hybrid frequency becomes equal to the integer of electronic cyclotron harmonics (see (1)). Fiber bursts are explained by completely different mechanism, the interaction of plasma waves with whistlers, propagating with the group velocity in the form of the periodic wave packets. The DPR- mechanism obtained wide recognition due to the most detailed developments of Zheleznyakov (1995) and subsequent series of articles and reviews of Zlotnik (Zlotnik et al. (2003). Zlotnik (2009), Zlotnik, (2010)).

Really, the theory seizes by the completeness of analytical analysis as well as the source model by its simplicity. The impression of the authenticity of model and operation of the mechanism is immediately created. This gives warrant to assume that the DPR- mechanism must work always, if there is the magnetic trap, in which the plasma density and the magnetic field strength fall down smoothly with the height in the corona with the different gradients. A question simultaneously arises, why then zebra- structure appears only irregularly (by per-second intervals) in the prolonged phenomena. In the images of SOHO/EIT then of TRACE and now also of SDO/AIA the magnetic flare loops remain almost constant during entire phenomenon. The repetitive maximums of continuous emission are suggestive of the presence of fast particles. But zebra-structure appears only irregularly during several seconds.

If we turn to laboratory plasma experiments, then we will be able to find no confirmation about the operation of DPR- mechanism. It is possible certainly to connect this with the difficulties of designing in the installation of the parameters of plasma, similar in the magnetic trap in the sun. Although the simulation of magnetic reconnection and excitation of ion-acoustic and whistler waves is long ago realized. The observations of the last years testify about the superfine structure of flare magnetic loops. In a number of works Aschwanden shown that in the thin loops the magnetic field weakly changes with the height (Aschwanden, 2004). In such loops it is not possible to expect even two DPR- levels, even with the most varied assumptions about the decrease in the density with the height. It is possible to remind that in some phenomena it was observed simultaneously more than 30 zebra stripes in the range 2.6 - 3.8 GHz. Thus, according to Figure 5.21 in the book Chernov (2011) it is shown that in any known models of coronal plasma it is impossible to obtain the large number of DPR- levels.

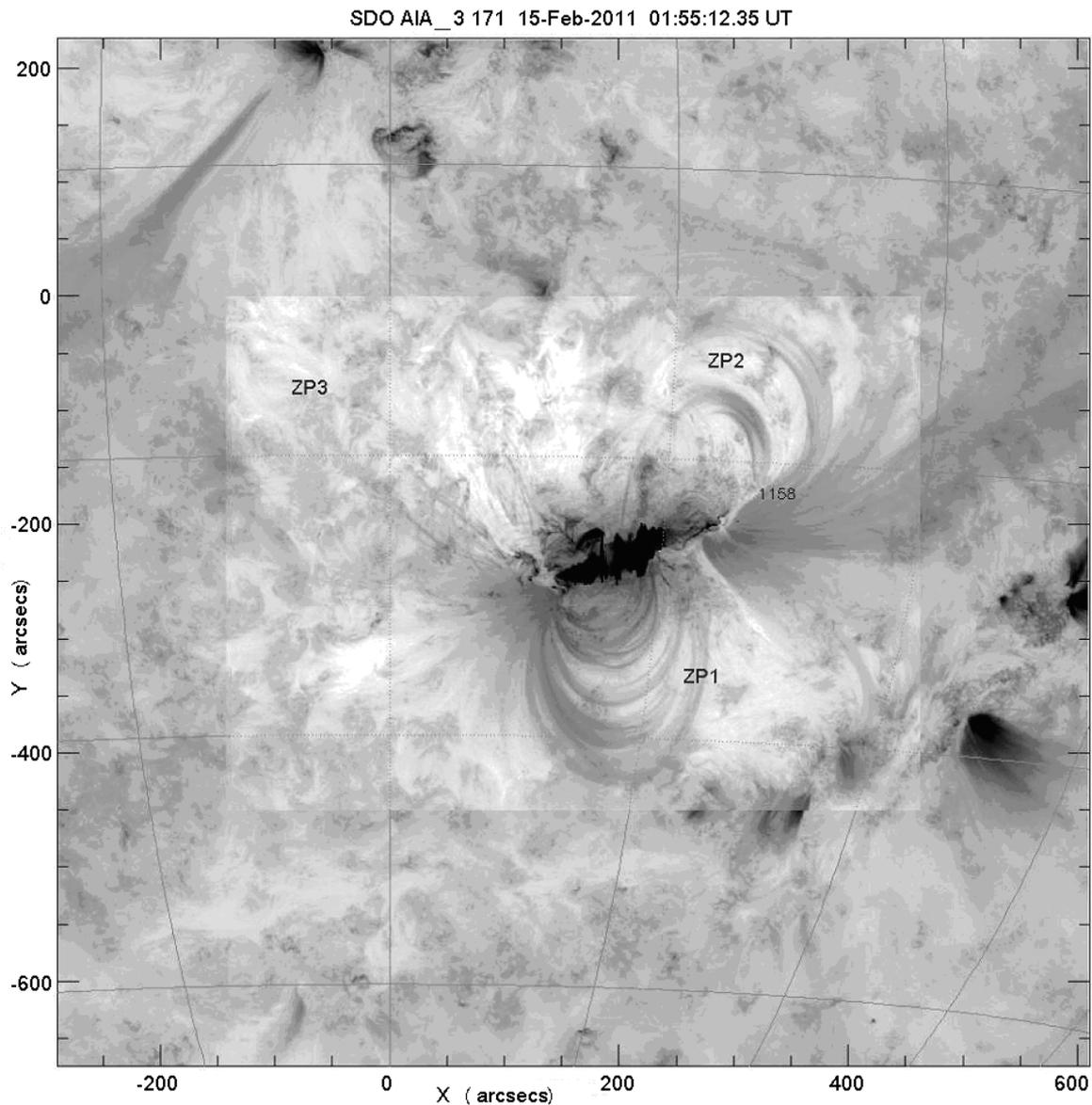

Figure 12. Position of possible sources of ZS (ZP1, ZP2, ZP3) defined on the directions of the simultaneous ejections in SDO/AIA 171 Å observations.

Nevertheless, the possibility of existence of the DPR levels in space plasmas has not been considered. Moreover, scale heights are taken arbitrarily in order to obtain several intersections of the curves of plasma frequency and electron cyclotron harmonics as, for example this is done in the recent paper by Chen et al (2011). It is suggested that the DPR- mechanism explains any phenomena with a zebra structure. Practically in all works discussing DPR- mechanism the presence of the large number of DPR- levels is considered factually. Only in Zlotnik et al (2003) is made an attempt to specifically identify the magnetic trap using magnetic field calculations above the AR for the 25 October 1994 event. However, the selection of main loop was made arbitrarily (in fact erroneously) without taking into account the fact that the particles critical for III type bursts and zebra- structure moved in the different directions (for greater detail, see Chernov et al (2005)).

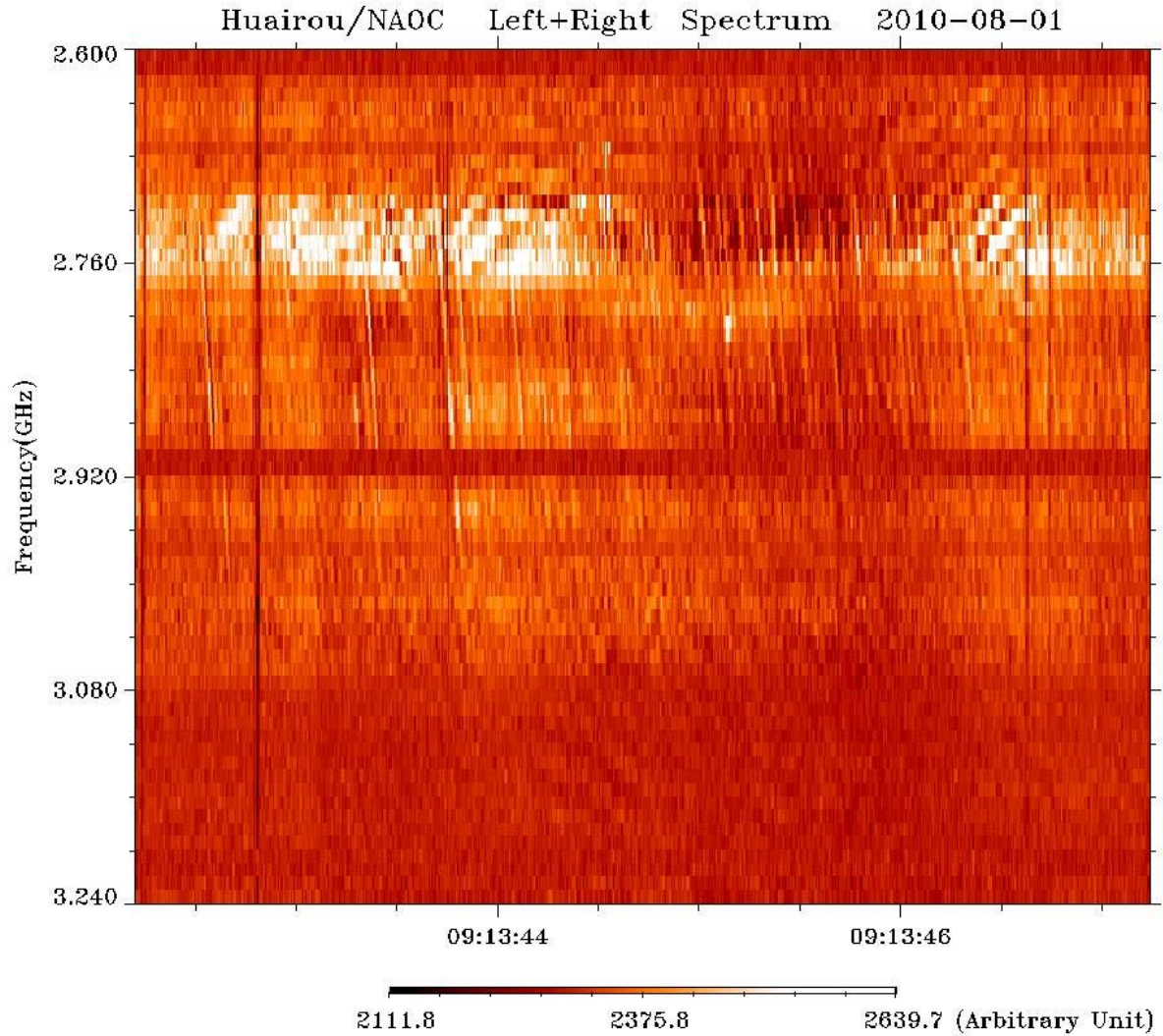

Figure 13. Two families of fiber bursts with opposite frequency drift as well as ZS at frequencies 3100 - 3230 MHz.

In several recent works the authors began to compare DPR- models and the interaction of plasma waves with whistlers. In this case an inaccuracy in the estimations of different parameters of whistlers were allowed. So, when considering the whistler model in Chen et al (2011), the projection of the group velocity of whistlers in the image plane is taken as the true value of the velocity, and the model was rejected allegedly without determining the true motion of sources.

In Tan et al. (2012) the estimation of the magnetic field strength using the frequency drift of fibers in the whistler model is made for the frequency of whistlers ($f_w$) relative to electron cyclotron frequency ($f_{ce}$) $x = f_w/f_{ce} = 0.01$, however, using an estimation of magnetic field based on the frequency separation of zebra stripes in the same frequency range it is $x = 0.25$. As a result in the latter case un underestimation of the field strength (by a factor 2) is obtained and the whistler model is rejected. The value $x$ is determined by the frequency, at which the value of the increment of whistlers is maximum. According to calculations of Chernov (2011) the value $x$ is between 0.1 and 0.01 depending on different parameters of the distribution function of fast particles. Therefore the field strength values in the whistler model must exceed estimations obtained in the DPR- model.

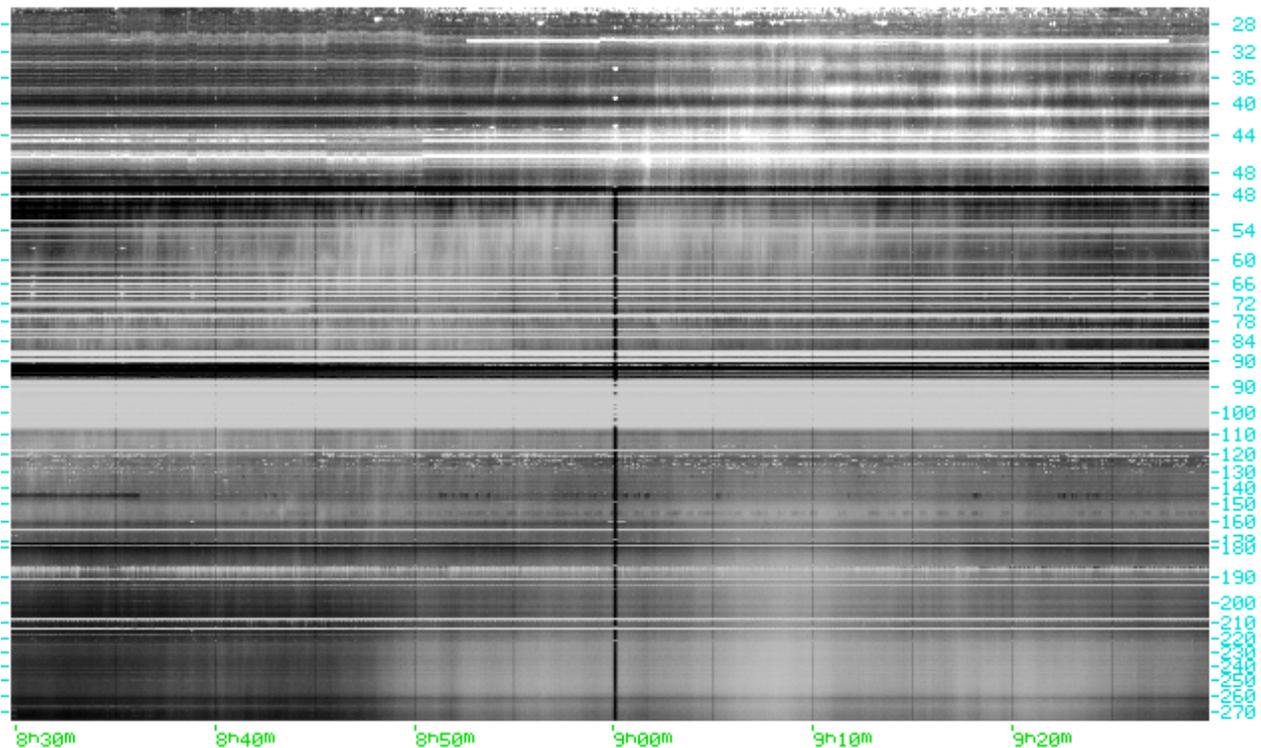

Figure 14. The type IV burst in the meter range (IZMIRAN, 25 - 270 MHz) on August 1, 2010. Against the continuum background only second pulsations with irregular period are distinguishable.

The slow drifting fibers in the type II burst (Figure 19) can be related with standing whistler wave packets before the shock front (Chernov, 1997). A magnetic trap is absent before the shock front, therefore in such a case the quasilinear interaction of whistlers with fast particles is also absent, and low frequency absorption does not form.

In the critical review of Zlotnik (2009), the advantages of the DPR model and the main failures of the model with whistlers are refined. The author asserts that the theory based on the DPR effect is the best-developed theory for ZP origin at meter-decimeter wavelengths at the present time. It explains in a natural way the fundamental ZP feature, namely, the harmonic structure (frequency spacing, numerous stripes, frequency drift, etc.) and gives a good fit for the observed radio spectrum peculiarities with quite reasonable parameters of the radiating electrons and coronal plasma. The statement that the theory based on whistlers is able to explain only a single stripe (e.g., a fiber burst) was made in Zlotnik (2009) without the correct ideas of whistler excitation and propagation in the solar corona.

Zlotnik uses the term "oscillation period" of whistlers connected with bounce motion of fast particles in the magnetic trap. Actually, the loss-cone particle distribution is formed as a result of several passages of the particles in the magnetic trap. Kuijpers (1975) explain the periodicity of fiber burst using this bounce period (~1 s). And if we have one injection of fast particles, whistlers (excited at normal cyclotron resonance) are propagating towards the particles (they disperse in space). Quasilinear effects thereby do not operate in normal resonance.

ZP is connected rather with whislers excited at anomalous resonance during long lasting injection. In such a case, waves and particles propagates in one direction, quasilinear effects begin operate and their role increases with increasing duration of injections. ZP is excited because the magnetic trap should be divided into zones of maximum amplification of whistlers, separated by intervals of whistler absorption (see in more details in Chernov ( 1990)). The bounce period does not interfere with this process, but it can be superimposed on ZP.

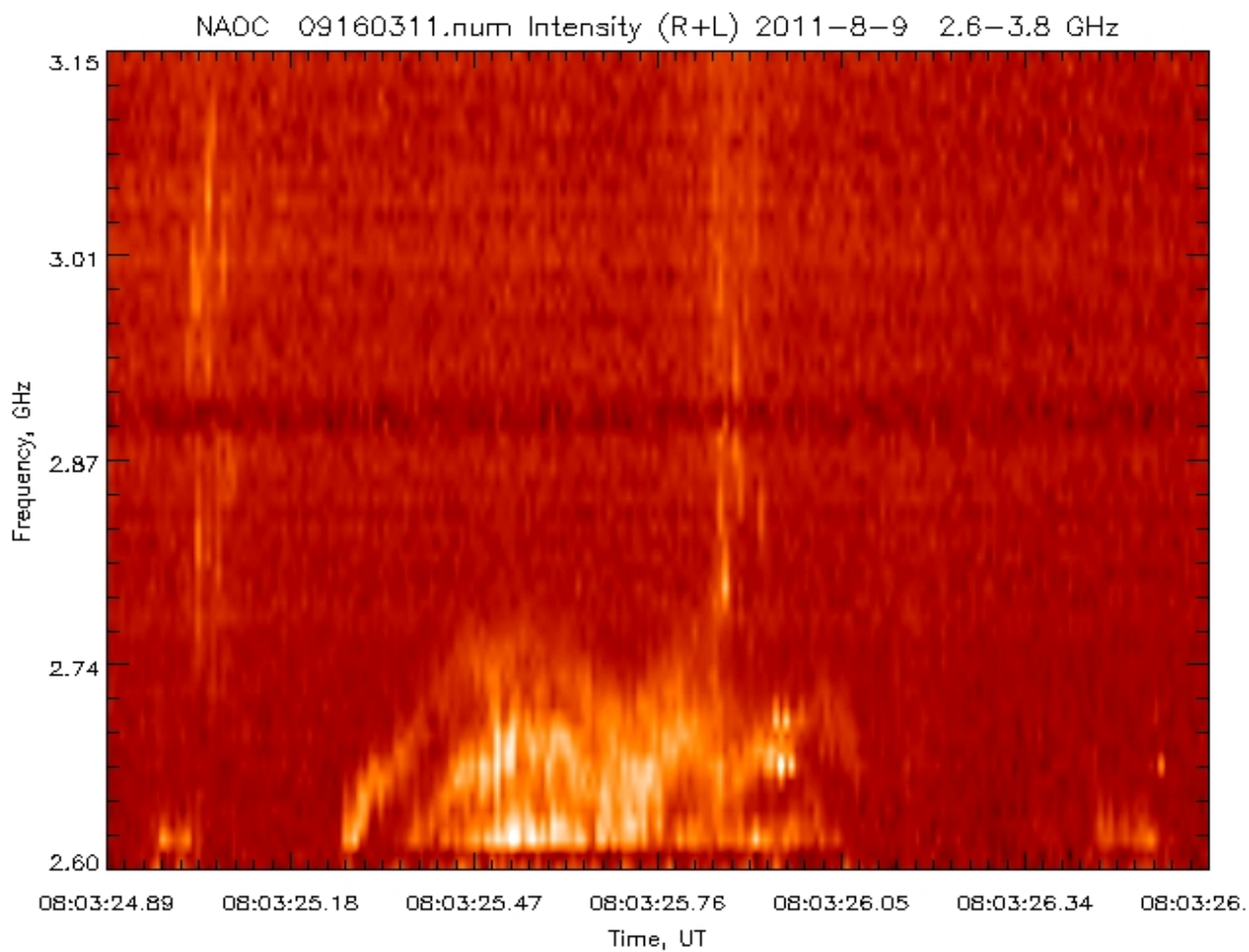

Figure 15. Three stripes of ZS in the LF edge of pulsations with millisecond superfine structure.

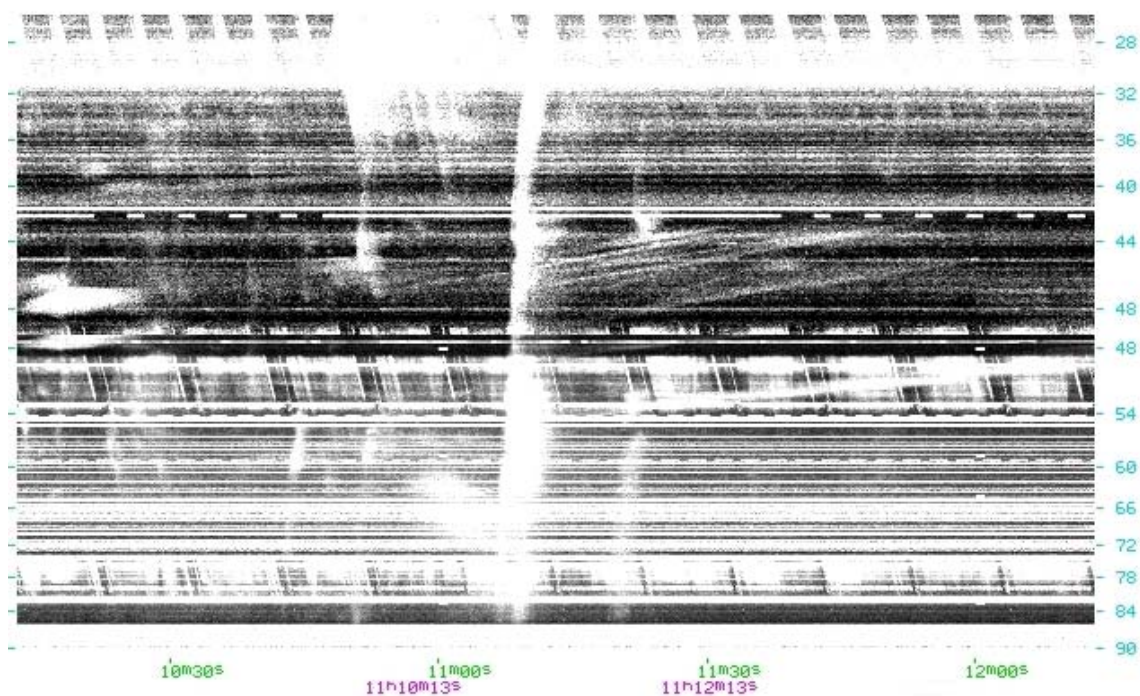

Figure 16a. Type II burst on 19 April 2012 in the range 25 – 90 MHz (IZMIRAN) consisting of slow drifting fibers.

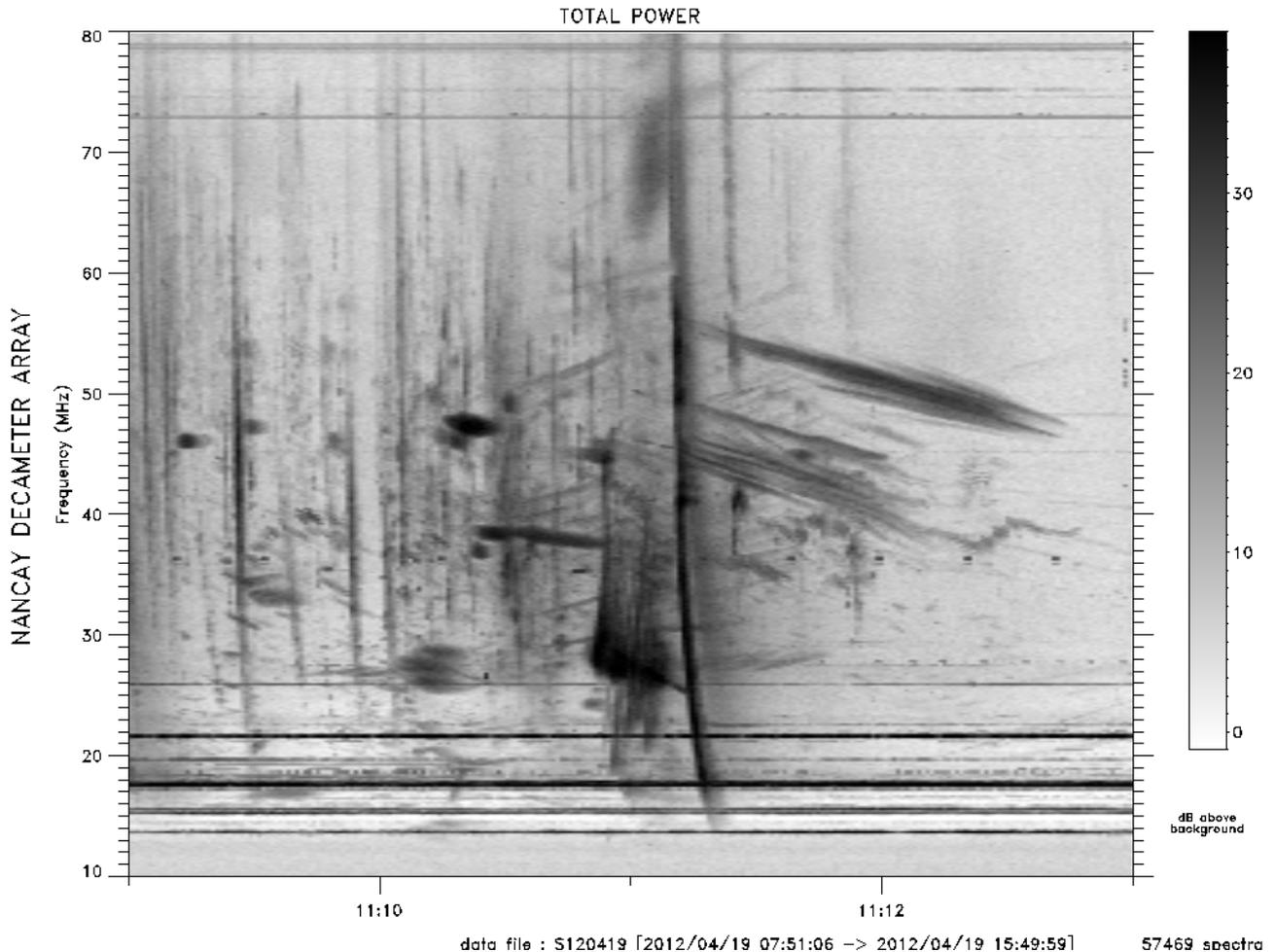

Figure 16b. Type II burst on 19 April 2012 in the range 10 – 70 MHz with *NANCAY Decametric Array* (courtesy Alain Lecacheux).

However, the whistler amplification length is always small (on the order of $\leq 10^8$ cm in comparison with the length of the magnetic trap being $>10^9$ cm) for any energy of fast particles (Breizman, 1987, Stepanov and Tsap, 1999). According to Gladd (1983), the growth rate of whistlers for relativistic energies of fast particles decreases slightly if the full relativistic dispersion is used. In this case, the whistlers are excited by anisotropic electron distributions due to anomalous Doppler cyclotron resonance.

Later, Tsang (1984) specified calculations of relativistic growth rates of whistlers with the loss-cone distribution function. It was shown that relativistic effects slightly reduce growth rates. According to Figure 8 in Tsang (1984), the relativistic growth rate is roughly five times smaller than the nonrelativistic growth rate. However, the relativistic growth rates increase with the perpedicular temperature of hot electrons $T_\perp$. According to Figure 5 in Tsang (1984), the growth rate increases about two times with the increase of electron energy from 100 to 350 keV, if other parameters of the hot electrons are fixed: loss-cone angle, ratio of gyrofrequency to plasma frequency, temperature anisotropy ($T_\perp/T_\parallel = 3$).

Thus, it has long been known that the whistlers can be excited by a relativistic beam with loss-cone anisotropy. Formula 13.4 in Breizman (1987), used as formula (29) in Chernov 2006) for evaluating the smallest possible relaxation length of beam, has no limitations in the value of energy of fast particles.

Critical comparison of models has been repeated in Zlotnik (2010), only with a new remark concerning the Manley-Rowe relation for the brightness temperature of electromagnetic radiation in result of coupling of Langmuir and whistler waves:

$$T_b = \frac{\omega T_l T_w}{\omega_l T_w + \omega_w T_l}. \qquad (2)$$

Zlotnik (2010) states that since $\omega_w \ll \omega_l$, in the denominator, only the first term remains and $T_b$ depends only on $T_l$, and $T_b \sim T_l$, i.e. the process does not depend on the level of whistler energy. However, Kuijpers (1975) (formula (32) in page 66) shown that the second term $\omega_w T_l$ should be $\gg \omega_l T_w$ due to $T_l \gg T_w$. Analogous conclusion was made by Fomichev and Fainshtein (1988) with more exact relation with three wave intensities (see also formula (11) in Chernov (2006)). Therefore the value of $T_b$ in the process $l + w \rightarrow t$ depends mainly on $T_w$.

Thus, our conclusion, that the entire magnetic trap can be divided into intermittent layers of whistler amplification and absorption remains valid for a broad energy range of fast particles.

In Zlotnik (2009) the main matter which is ignored is that the model involves quasilinear interactions of whistlers with fast particles, allowing one to explain all the fine effects of the ZP dynamics, mainly the superfine structure of ZP stripes and the oscillating frequency drift of the stripes which occurs synchronously with the spatial drift of radio sources.

Continuous discussions stimulate the developments of new models. Treumann, Nakamura, and Baumjohann (2011) a new mechanism for ZP based on an ion-cyclotron maser mechanism. Thanks to the special delta-shaped distribution function of the accelerated ions, the ion-cyclotron maser generates a number of electromagnetic ion-cyclotron harmonics which modulate the electron maser emission. A part of the accelerated relativistic protons pass along the magnetic field across the trapped loss-cone electron distribution. The modulation of the loss-cone will necessarily lead to a modulation of the electron cyclotron maser. Locally this produces the typical "Zebra" emission/absorption bands. However this mechanism can only work in strong magnetic field, when $f_{pe}/f_{ce} < 1$.

Karlicky et al. (2013) continued the development of the model of Kuznetsov (2006) for fiber bursts: fiber bursts can be explained by a propagating fast sausage magnetoacoustic wave train. Then Karlicky (2013) extended a similar model for ZP: the magnetoacoustic waves with density variations modulate the radio continua, and this modulation generates zebra effects. It should be noted that close model was examined earlier in three works by Laptuhov and Chernov (2006; 2009; 2012).

Yurovsky (2011) continued a series of works about the formation ZP due to the refraction (interference mechanism) of rays on the heterogeneities in the corona. However the author doesn't refer to the previous works (Ledenev, Yan, and Fu 2006), and carried out a simplified simulation of ZP stripes which matches with those observed.

## 4. CONCLUSIONS

The model involving whistlers explains many special features the zebra- structure:
- the oscillatory frequency drift and the frequency splitting of stripes,
- a change in the spatial drift of radio source synchronously with the frequency drift of stripes in the spectrum,
- the millisecond superfine structure of stripes.

The simultaneous appearances of fibers and zebra-structure shown here and the smooth transition of zebra-stripes into the fibers and back testify in favor of the united mechanism of the formation of different drifting stripes in emission and absorption within the framework of the interaction of plasma waves with whistlers, taking into account quasi-linear interaction of whistlers with the fast particles and with the ion-acoustic waves.

The presence of ion-acoustic waves can be considered justified in the source in the region of magnetic reconnection with the outgoing shock fronts. The propagating ion-acoustic waves can serve as the natural heterogeneities, through which electromagnetic waves pass that can lead to shaping of the additional stripes of transparency and opacity in the spectrum. All components of this mechanism are examined in Laptukhov, Chernov (2009, 2012). It should be noted that the relative significance of these possible mechanisms remains uncertain.

Simultaneous or consecutive appearance the zebra-structure in different frequency ranges is obviously related to the dynamics of flare processes, which is supported by the phenomena examined here. For a comparative analysis of observations of zebra-structure and fiber bursts and different theoretical models we refer to the review by Chernov (2012).

Acknowledgements

We are grateful to the Nobeyama, SOHO (LASCO/EIT), and SDO/AIA teams for operating the instruments and performing the basic data reduction, and especially for their open data policies.
This work was supported by a Chinese Academy of Sciences Visiting Professorship for Senior International Scientists, grant No. **2011T1J20**), and partially supported by the Russian Foundation of Basic Research (RFBR), grant Nos. 14–02–00367. The National Basic Research Program of the Ministry of Science and Technology of China (Grant No. 2006CB806301) and CAS-NSFC Key Project (Grant No. 10778605), provided support for the Chinese authors.